\documentclass[11pt]{article}
\pdfoutput=1
\usepackage{amsmath,epsf,amssymb,latexsym,enumerate,cite,shadow,array,color}
\usepackage{slashed}
\usepackage{graphicx,wasysym}

\DeclareFontFamily{U}{rsf}{} \DeclareFontShape{U}{rsf}{m}{n}{
  <5> <6> rsfs5 <7> <8> <9> rsfs7 <10-> rsfs10}{}
\DeclareMathAlphabet\Scr{U}{rsf}{m}{n} \makeatletter
\@addtoreset{equation}{section} \makeatother

\def\be{\begin{equation}}
\def\ee{\end{equation}}
\def\ba{\begin{array}}
\def\ea{\end{array}}

\newcommand{\bea}{\begin{eqnarray}}
\newcommand{\eea}{\end{eqnarray}}

\def\K{K{\"a}hler}
\newcommand{\ft}[2]{{\textstyle\frac{#1}{#2}}}

\def\rme{{\rm e}}
\usepackage{ifpdf}
\ifx\pdfoutput\undefined
   \pdffalse
\else
   \pdfoutput=1
   \pdftrue
  \usepackage[pdftex]{hyperref}
\newcommand{\rf}[1]{(\ref{#1})}

\def\K{K{\"a}hler}

\def\F{Friedmann}

\begin{document}

\begin{titlepage}

\

\

\begin{center}
{\LARGE \textbf{ Hidden Superconformal Symmetry\\
 of the Cosmological Evolution 
\vskip 2cm }}

{\bf Renata Kallosh and  Andrei Linde} \\

\

{\sl Department of Physics, Stanford University, Stanford, California
94305 USA}
\end{center}
\vskip 2.5 cm

\begin{abstract}

In the superconformal formulation of supergravity, the standard supergravity action appears as a result of spontaneous symmetry breaking when the conformal compensator scalar field, the conformon, acquires a nonzero value, giving rise to the Planck mass. After that, many symmetries of the original theory become well hidden, and therefore they are often ignored. However, recent developments demonstrated that superconformal invariance is more than just a tool: it plays an important role in generalizing previously existing formulations of supergravity and developing new classes of inflationary models.  In this paper we describe hidden superconformal symmetry of the cosmological evolution. In this formulation, inflation can be equivalently described as the conformon instability, and creation of the universe `from nothing' can be interpreted as spontaneous symmetry breaking due to emergence of a classical conformon field. We develop a general formalism that allows to describe the cosmological evolution simultaneously with the evolution of the conformon. We find a set of gauge invariant physical observables, including the superconformally invariant generalizations of the square of the Weyl tensor, which are necessary for invariant description of the cosmological singularities.

\end{abstract}

\vspace{24pt}
\end{titlepage}

 \tableofcontents

\newpage
\parskip 7pt

\section{Introduction}

The standard approach to cosmological evolution is based on the Einstein theory of gravity. The gravitational constant in this theory is indeed a constant, $G = (8\pi M_{p})^{-2}$. Since it is a constant, it is customary to simply take $M_{p} = 1$ in all equations. In the standard approach to supergravity, one can also take $M_{p} = 1$. However, in the superconformal formulation of supergravity, which is one of the most powerful tools used since the very early days of this theory  \cite{Siegel:1977hn,Kaku:1978ea,Cremmer:1978hn,Freedman:2012zz}, supergravity possesses an additional set of fields and symmetries. Planck mass becomes a constant only after gauge fixing of these extra symmetries. Just like in the theory of spontaneous symmetry breaking in the Higgs model, the original symmetry does not disappear after the symmetry breaking and/or gauge fixing: One can either use the unitary gauge, where the physical contents of the theory are manifest, or other gauges where the calculations can be easier; all physical results do not depend on this choice. Similarly, the original symmetries of the superconformal theory are still present in the standard formulation of supergravity with $M_{p} = 1$, but they are well hidden and therefore often forgotten and rarely used.

Recent cosmological developments forced us to return back to the basics, reformulate the superconformal formulation of supergravity in a way especially suitable for cosmological applications \cite{Kallosh:2000ve,Ferrara:2010yw}, and apply it to the development of a new class of inflationary theories with interesting universality properties \cite{Kallosh:2013pby,Kallosh:2013lkr,Kallosh:2013hoa,Ferrara:2013rsa,Kallosh:2013daa,Kallosh:2013tua,Kallosh:2013yoa}. Each new step of this way suggested that the superconformal approach is not just a tool for the development of supergravity, but a convenient framework which deserves full attention on its own merits. In this paper we will make a step towards a maximally symmetric representation of the cosmological evolution, using all symmetries of the superconformal theory in a democratic way. %In this approach, the superconformal theory 

  One starts with the model which has no dimensionful parameters but has a local 
 Weyl symmetry. In particular the curvature term in generic superconformal theories always has a coupling to scalars,    of the form $-{1\over 6} {\cal N}(X^I, \bar X^{\bar J}) \, R$, where $X^I, \bar X^{\bar J}$ are complex scalars.  Here $ {\cal N}(X^I, \bar X^{\bar J})$ is the \K\, potential of the embedding manifold, including the negative signature conformon field $X^0$.  When Weyl symmetry is spontaneously broken, for example, by requiring that the \K\, potential of the embedding manifold is constant,  ${\cal N}(X^I(x), \bar X^{\bar J}(x)) = 3  M_{Pl}^2$,  one recovers the standard supergravity and general relativity in the Einstein frame \cite{Kallosh:2000ve}. However, one can also use a different Weyl gauge choice,  ${\cal N}(X^I(x), \bar X^{\bar J}(x)) = 3 M_{Pl}^2 e^{-{\cal K}(z, \bar z)/ 3 M_{Pl}^2} $, where ${\cal K}(z, \bar z)$ is the \K\, potential of supergravity with only physical scalars. This approach to cosmology was developed and used for constructing a supersymmetric version of the Higgs inflation where the Jordan frame with some non-minimal couplings of scalar to gravity plays an important role \cite{Ferrara:2010yw}.  

 Yet another possibility is to make an investigation without fixing any gauge for as long as possible, and then turn to the Einstein frame only at the very end of the calculations, when the comparison with observations is made. This possibility proposed in  \cite{Kallosh:2000ve} may be quite appropriate for investigation of physical processes in the Friedmann universe. Indeed, the FLRW metric is conformally flat. Therefore in the formulation where the conformal symmetry is not broken by gauge fixing, one can reduce investigation of physical processes in an expanding Friedmann universe to a study of processes in a flat Minkowski space, and then fix the gauge (and the Planck mass) at the very end of the investigation. Previously this method was often used for investigation of ultra-relativistic particles in the early universe, but now a generalized version of this method becomes available for investigation of all supergravity-based models, by using their superconformal formulation.

 By paying attention to the full symmetry of the superconformal theory, one may develop a different attitude to what is natural and what is unnatural in particle physics and cosmology. One may find many unexpected links between various cosmological theories which previously could seem entirely unrelated. Recent developments have demonstrated that hidden (spontaneously broken) superconformal invariance of supergravity allows to  generalize previously existing versions of cosmological models in supergravity and to develop new classes of inflationary models, which lead to an attractor behavior of physical observables \cite{Kallosh:2013pby,Kallosh:2013lkr,Kallosh:2013hoa,Ferrara:2013rsa,Kallosh:2013daa,Kallosh:2013tua,Kallosh:2013yoa}, in agreement with the recent cosmological data from WMAP9  \cite{Hinshaw:2012aka} and Planck 2013 \cite{Ade:2013rta}. One of the key features of a broad class of inflationary theories developed in \cite{Kallosh:2013hoa} is an $SO(1,1)$ deformed symmetry between the conformon and the inflaton. It would be very hard to describe this symmetry after the Weyl symmetry is spontaneously broken, or, equivalently, gauge-fixed. It is a feature of the Weyl invariant theory. Starting with the Weyl symmetry and the $SO(1,1)$ symmetry, and then deforming $SO(1,1)$ and fixing the gauge, is reminiscent of using all symmetries of special theory of relativity and then returning to the laboratory reference frame at the end of the calculations.

We will show in this paper that hidden conformal and superconformal symmetries may provide a useful tool for description of the cosmological evolution, in general.  Here we focus on two applications.
First, we will study the approach to inflation with account of the fact that the FLRW universe is conformally flat and deviations from FLRW are small during inflation. Instead of $n$ scalars evolving in 
conformally flat FLRW universe with the time-dependent scale factor $a$ we will have a set-up where $n+1$ scalars, including an extra scalar, the  conformon, evolve in a flat space with $a=1$. New aspects of interpretation of cosmological evolution in the context of models with spontaneously broken conformal symmetries will be studied. For example, we will find out that inflation can be equivalently described as the conformon instability.

We will also investigate the evolution in the opposite direction, towards the cosmological singularity in this conformal setting. In the standard Einstein theory of relativity the general cosmological solution of classical Einstein equations has a Big Bang  time singularity. It manifests itself in the fact that the density of matter and invariants of the Riemann curvature tensor blow up. However, if we would start with conformally invariant gravity and break conformal symmetry spontaneously, would it be possible to avoid the singularity?   To address this issue we will look for the geometric invariants which are Weyl invariant as well as invariant under the change of coordinates. They will help to distinguish the true singularities from the ones which can be avoided by a choice of coordinates or  conformal geometries.

In this paper we will investigate the superconformal approach to the cosmological evolution in the theory of chiral multiplets; a generalization for vector or tensor multiplets can be performed following the lines of \cite{Ferrara:2013rsa}.

In Section 2 we start with a toy conformally invariant model describing an exponentially expanding de Sitter universe. We will discuss two equivalent descriptions of this processes, related to each other by a gauge transformation. In one of these descriptions, we deal with an exponentially expanding de Sitter space with a positive cosmological constant. In the second formulation, the universe does not expand at all, it is a flat Minkowski space containing an exponentially growing conformon field. We explain that this is a general result, which implies that the cosmological expansion in supergravity can be equivalently described either by expansion of the scale factor of the universe, or by the growth of the conformon field. 

In Section 3 we continue discussion of the toy model studied in Section 2. We develop canonical formalism describing the cosmological expansion of the universe in this model, as well as the dual description of the cosmological evolution in terms of the growth of the conformon field. In addition, we derive the Wheeler-DeWitt equation for the wave function of the universe, and show that it has the same functional form independently of whether we consider the wave function of the universe $\Psi(a)$ depending on the scale factor, or the wave function of the universe $\Psi(\chi)$ depending on the conformon field.

 In Section 4 we study cosmological evolution in a generic superconformal theory. First, we explain that a particular ansatz for the solutions of equations of motion of such models, corresponding to the so-called Einstein frame conformal gauge,  leads to standard Einstein equations of motion for the bosonic part of general superconformal theory. Secondly, we show how to study these models in an arbitrary Jordan frame. Finally, we use the general ansatz for the metric which allows to derive equations of motion of the conformal theory without making a choice of any particular Weyl gauge, these are conformally covariant equations.

In Section 5 we develop a generic superconformal framework for conformally flat \F\, universe. We focus on a  Weyl gauge which makes the scale factor of the universe time-independent. The resulting evolution equations are  the geodesic equations for scalars in flat space  in embedding \K\, geometry. These are equivalent to standard general relativity equations, including the Friedmann equation for the scale factor. The Hamiltonian formalism has a nice interpretation in this formalism.

In Section 6 we provide a framework for physical observables in Weyl  invariant models. We present Weyl  curvature invariants which may be used in these models to infer the gauge-independent features of the models. These invariants will represent the true singularities which are not removable  by the change of the coordinates as well as by the change of the Weyl gauge.

In Section 7 we give a summary of our results.

%%%%%%%%%%%%%%%%

\section{de Sitter space and a runaway solution for the conformon}\label{toy}

In this section we will discuss a simple toy model which will be useful for explaining the basic concepts to be discussed in the paper.

\subsection{A toy model: Conformally invariant theory of the cosmological constant}

Consider a simple theory of a scalar field $\xi$ interacting with gravity as follows:
\begin{equation}
\mathcal{L} = \sqrt{-{g}}\left[{1\over 2}\partial_{\mu}\chi \partial_{\nu}\chi \, g^{\mu\nu}  +{ \chi^2\over 12}  R({g}) -{\lambda\over 4}\chi^4\right]\,.
\label{toy2a}
\end{equation}
This
theory is locally conformal invariant under the following
transformations: 
\be\label{conf2}
g_{\mu\nu} \to \rme^{-2\sigma(x)} g_{\mu\nu}\, , \qquad
\tilde\chi \to  \rme^{\sigma(x)} \chi\ . 
\ee
The field $\chi(x)$ is referred to as  a conformal compensator, which we  call `conformon'  \cite{Kallosh:2000ve}.   It has negative sigh kinetic term, but this is not a problem because it
can be removed from the theory by fixing the gauge symmetry
(\ref{conf2}),  for example by taking a gauge $\chi =\sqrt 6 M_{p}$, where $M_{p}$ is the Planck mass.  This gauge fixing can be interpreted as a spontaneous breaking of conformal invariance due to existence of a classical field $\chi =\sqrt 6 M_{p}$. We will keep $M_{p} = 1$ throughout the paper, but it is useful to remember the relation between the Planck mass and the conformon field in this gauge.

\subsubsection{$\chi = \sqrt 6$ conformal gauge}
After fixing $\chi = \sqrt 6$, the kinetic term of the scalar field disappears, the term ${ \chi^2\over 12}  R({g})$ becomes the standard Einstein action, and the  term ${\lambda\over 4}\chi^4$ becomes a cosmological constant $\Lambda = {9}\lambda$:
\be
\mathcal{L}=  \sqrt{-{g}}\,\left[{  R({g})\over 2}-{9\lambda}\right]\, .
\label{toy2b}
\ee
This theory has a simple de Sitter solution 
with metric
\be \label{friedmann}
ds^{2} = -dt^{2} + a^{2}(t) d\vec x^{2} \ ,
\ee
where 
\be\label{dssol}
a(t) = e^{Ht}  = e^{\sqrt{\Lambda/3}\ t}  = e^{\sqrt{3\lambda}\ t} \ ,
\ee
and
\be
H = \sqrt{\Lambda/3} = \sqrt{3\lambda} \ .
\ee
One can also make a change of variables $d\eta = {dt}/a(t)$ and write the metric (\ref{friedmann}) in a conformally flat form,
\be \label{friedmann2}
ds^{2} =a^{2}(\eta)[-d\eta^{2} + d\vec x^{2}] \ .
\ee
For de Sitter space with $a(t) = e^{Ht}$ this yields
\be\label{eta22}
\eta = -H^{{-1}} e^{{-Ht}} = - {1\over \sqrt{3\lambda}}e^{-\sqrt{3\lambda}\ t} \ ,
\ee
and therefore
\be \label{friedmann22}
ds^{2} ={1\over H^{2}\eta^{2}}(-d\eta^{2} +  d\vec x^{2}) ={1\over 3\lambda\eta^{2}}(-d\eta^{2} +  d\vec x^{2}) \ .
\ee
Here we made a normalization $\eta = -H^{{-1}}$ for $t = 0$. Note that $\eta$ runs from $-\infty$ to $0$ when $t$ runs from $-\infty$ to $+\infty$.

\subsubsection{$a=1$ conformal gauge}

Instead of the gauge $\xi = \sqrt 6$, one may also use the gauge $a = 1$.
In this gauge,  the metric is flat in conformal time,
\be\label{confflat}
ds^{2} =-d\eta^{2} +  d\vec x^{2} \ ,
\ee
and the theory describes the scalar field $\xi$ in flat Minkowski space. The action becomes
\begin{equation}
\mathcal{L} = {1\over 2}\partial_{\mu}\chi \partial_{\nu}\chi \, \eta^{\mu\nu}   -{\lambda\over 4}\chi^4\,.
\label{toya=1}
\end{equation}
Equation of motion for the field $\chi$ in Minkowski space is
\be
 \chi'' =\lambda \chi^{3} \ .
\ee
Here $ \chi''   = {d^{2} \chi\over d \eta^{2}}$. Note that because of the ``wrong'' sign of the kinetic term of the curvaton field, its equation of motion is the same as of the normal field with a negative potential $-{\lambda\over 4} \chi^4$. Therefore the conformon field experiences an instability, falling down in its potential unbounded from below. This equation has a general solution (up to a time redefinition $\eta \to \eta -\eta_{0}$), such that $  \chi \to  +\infty$ for $\eta$ growing from $-\infty$ to $0$:
\be\label{chieq}
 \chi = -{\sqrt 2\over \sqrt \lambda \eta} \ .
\ee

\subsubsection{Relation between gauges} 
To compare our result (\ref{chieq})   to the results obtained  in the gauge $\chi=\sqrt 6$, one can  use the conformal transformation \rf{conf2} with $e^\sigma= -\sqrt {3\lambda} \eta$:
\be\label{chiexpand}
 \chi = -{\sqrt 2\over \sqrt \lambda \eta}\, , \qquad \Rightarrow  \qquad  \chi =  {\sqrt 2\over \sqrt \lambda \eta} \,  \sqrt {3\lambda} \eta=\sqrt 6 \ .
\ee
The flat metric of the $a=1$ gauge becomes  $ \rme^{-2\sigma(x)} \eta_{\mu\nu}$,
\be\label{confflat3}
ds^{2} = {1\over 3\lambda \eta^{2}} (-d\eta^{2} + dx^{2}) \ ,
\ee
which coincides with \rf{friedmann22}, as it should. Finally, form this metric one can recover the usual Friedmann metric by requiring that $\eta < 0$, and therefore from $a(\eta) d\eta = {dt}$ one finds that 
\be\label{eta}
\eta = -H^{{-1}} e^{{-Ht}} = - {1\over \sqrt{3\lambda}}e^{-\sqrt{3\lambda}\ t} \ ,
\ee
which brings back the dS solution  (\ref{dssol}), (\ref{eta22}), which we earlier obtained by the standard method.

\subsection{Interpretation and consequences: Inflation as the conformon instability}

Let us say few words about interpretation of our result, which will turn out to be much more general than the simple model discussed so far. In order to do it, let us express the value of the conformon field $\chi$ in a non-expanding Minkowski space \rf{chiexpand} in terms of time $t$ in the Friedmann universe:
\be\label{expconformon}
\chi = -{\sqrt 2\over \sqrt \lambda \eta} = {\sqrt 6}\, e^{{H\, t}} = {\sqrt 6} M_{p}\, e^{{H\, t}} \ .
\ee
Note that since the theory is locally conformally invariant, one can always ``freeze'' the evolution of the conformon field at any moment $t$, and allow the scale factor to evolve starting from this moment, by making a proper conformal transformation, or choosing an appropriate gauge.  The corresponding wavelength, corresponding to the effective Planck length, decreases as $e^{{-H\, t}}$. Thus, Minkowski space does seem exponentially expanding if its size is measured in units of the exponentially contracting Planck length. This is a general result, which is applicable to any kind of uniform cosmological evolution, including inflation. In this context, exponential growth of space during inflation \rf{dssol} is directly related (equivalent) to the exponential growth of the conformon field in Minkowski space (\ref{expconformon}).

In order to understand this general result, which is going to be valid for all models studied in this paper, it is sufficient to look at the equation (\ref{conf2}). In the standard investigation of the cosmological evolution, one goes to what can be called the Einstein frame gauge, fixes the conformon $\chi = \sqrt 6$ (or, more generally, the Planck mass), and investigates the evolution of the scale factor $a$, as measured in the Planck length units.  However, one can equally well work in the gauge where the scale factor is fixed. The transition from one gauge to another is achieved by conformal transformation (\ref{conf2}), which absorbs expansion of the universe in terms of its scale factor $a(t)$ and converts it into the exactly equal time-dependent factor describing the growth of the conformon field.

In application to inflation, this means that one can equally well describe it as the exponentially fast expansion of the scale factor, or as the equally fast growth of the conformon field, obeying {\it the same Einstein equations as the scale factor, up to a trivial rescaling}. Alternatively, one can work in the original conformally invariant setting, without fixing the gauge, and study evolution of all fields while preserving the original conformal invariance and enjoying simplifications provided by conformal flatness of the Friedmann universe. Then in the end of the calculations one can re-formulate all results in terms of the Einstein frame gauge where the Planck mass is fixed.

\section{Canonical formalism in a conformally flat Friedmann universe}
Following \cite{DeWitt:1967yk} we take an ansatz for the metric\footnote{In this part of the paper we use the standard notation for the lapse function $N(t)$, which should not be confused with the number of e-folds $N$ during inflation.}
\be 
ds^{2} = -N^2(t)dt^{2} + a^{2}(t) d\vec x^{2} \ .
\label{deW}
\ee
Here $N(t)$ is a lapse function which is a useful auxiliary variable in canonical gravity formalism. 
In addition to the choice of the metric in \rf{deW} we assume that  $\chi$  depends only on $t$. Without fixing local conformal symmetry we find the following Lagrangian:
\begin{equation}
\mathcal{L} = N(t)  a(t)^3 \left[-{1\over 2 N^{2}(t)} \dot \chi ^2   +{ \chi^2\over 12}  R({g}) -{\lambda\over 4}\chi^4\right]\, ,
\label{toydeW}
\end{equation}
where for the metric (\ref{deW})
\be
R(g)= {6(-a \dot a \dot N + N ( \dot a^2 + a \ddot a))\over a^2 N^3} \ .
\label{R}\ee

\subsection{Gauge $\chi=\sqrt 6$}

 In the conformal  gauge $\chi=\sqrt 6$ we find 
\begin{equation}
\mathcal{L} =   a^3 \Big ( -{  3  \dot a^2  \over  a^2 N} 
 -  9 \lambda N  \Big ) \, ,
\label{toydeW1}
\end{equation}
up to the boundary term which can be ignored for the description of a compact universe, or canceled by adding Gibbons-Hawking terms to the action.

The canonical momenta are $\pi_{a} ={\partial L \over\partial a} = -{6\dot a a\over N}$ and $\pi_{N} ={\partial L \over\partial N} = 0$. The Hamiltonian is 
\be
{\mathcal H} = -{N\over 12 a}\pi_{a}^{2} + 9\lambda N a^3   
\ee
Equation $\pi_{N} ={\partial L \over\partial N} = 0$ leads to the Hamiltonian constraint \cite{Arnowitt:1959ah}
\be\label{vanishingH}
{\mathcal H} = -{1\over 12 a}\pi_{a}^{2} + 9\lambda  a^3   =0 
\ee
with the choice of time variable corresponding to the lapse function  $N(t) = 1$. It is equivalent to the standard Einstein equation for the universe with the cosmological constant $V = 9\lambda$:
\be
H^{2} = \Big ( {   \dot a  \over  a } \Big)^2 
 =  {V\over 3}\ .
\ee
  A similar equation can be written for a closed or an open universe, but here we limit ourselves to the simplest case of a flat Friedmann universe.

One can use these results to write the Wheeler-DeWitt equation for the wave function of the universe. Usually it is done for a closed universe with scale factor $a$,
\be\label{WDWa}
\left[{1\over 24\pi^{2}} {d^{2}\over d a^{2}} - 6\pi^{2} a^{2} + 2\pi^{2} a^{4} V\right] \Psi(a) = 0 \ ,
\ee
which describes quantum creation of the universe \cite{Vilenkin:1982de,Hartle:1983ai}. However, according to \cite{Linde:1983cm,Rubakov:1984bh,Zeldovich:1984vk,Vilenkin:1984wp}, the probability of this process is exponentially suppressed by $\exp\left(-{24\pi^{2}\over V}\right)$. Therefore here we will consider the Wheeler-DeWitt equation which may describe the probability of quantum creation of a compact flat universe, which is a box of size $a$ with identified opposite sides, i.e. a torus. Since compactification of extra dimensions is part and parcel of string theory, it is natural to extend this idea to all spatial dimensions. In this case, the Wheeler-DeWitt equation looks as follows:
\be
\left[{d^{2}\over da^{2}} + 12 a^{4} V\right]\Psi(a) = 0 \ ,
\ee
or, equivalently,
\be\label{lambdaWDW}
\left[{d^{2}\over da^{2}} + 108\lambda a^{4}\right]\Psi(a) = 0 \ .
\ee
One could expect that creation of a topologically nontrivial universe should be even stronger suppressed, but in fact an opposite is true. The solution of this equation is not exponentially suppressed because it does not involve any tunneling, see a discussion of this issue in \cite{Zeldovich:1984vk,Coule:1999wg,Linde:2004nz}.

Now we will compare these results with the similar results in the gauge $a= 1$.

\subsection{Gauge $a= 1$}

 In the conformal  gauge $a(t)= 1$ 
where $R=0$ (for arbitrary lapse function $N(t)$, as one can see from \rf{R}) the Lagrangian  is
  \begin{equation}
\mathcal{L} =    -{1\over 2 N(t)} \dot \chi ^2   -{\lambda\over 4} N(t) \chi^4\,.
\label{toydeW2}
\end{equation}
In this gauge it does not depend on the lapse velocity $\dot N$.
Equation for $N(t)$ is
\be
{\partial {\mathcal L}\over \partial N(t)}= {1\over 2 N^2} \dot \chi ^2   -{\lambda\over 4}  \chi^4=0\, \ .
\ee
For the choice of a conformal time variable $\eta$ above we find a constraint
\be
{1\over 2 } ( \chi' )^2   -{\lambda\over 4}  \chi^4=0\, ,
\ee
which $ \chi = -{\sqrt 2\over \sqrt \lambda \eta}$ indeed satisfies.

From the Lagrangian in the gauge $a(t)= 1$, where $R=0$, we define the canonical variables and constraints
\be
\pi_\chi = -{1\over N} \dot \chi \, \qquad \pi_N = 0\ ,
\ee
and \be
H= \int (\pi_N \dot N + \pi_\chi \dot \chi )  d^3 x -{\cal L} = \int d^3 x (\pi_N \dot N + N {\cal H}) \ ,
\ee
where in our example
\be
{\mathcal H}= - {1\over 2} \pi_\chi^2+ {\lambda\over 4} \chi^4 \ .
\label{Hconf}\ee
The primary constraint $\pi_N=0 $ is associated with the {\it secondary or dynamical constraint}, the Hamiltonian constraint $\mathcal{H}=0$, which is fully analogous to \rf{vanishingH}. 
It is the ``kinetic energy'' of the scale factor of the universe that gives a rather unusual negative contribution to the total Hamiltonian in \rf{vanishingH}, which exactly cancels the positive contribution of the vacuum energy and allows ${\mathcal H}$ to vanish. Meanwhile, the negative contribution of the conformon energy to the Hamiltonian \rf{Hconf}  in the gauge $a = 1$ is a simple consequence of the fact that the conformon has a negative signature metric in the moduli space of scalars, by construction, there is nothing mysterious about it.

For a topologically nontrivial flat universe considered in the previous subsection, but now having constant size $a = 1$, the Wheeler-DeWitt equation becomes
\be\label{lambdaWDWconf}
\left[{d^{2}\over d\chi^{2}} + {\lambda\over 2} \chi^{4}\right]\Psi(\chi) = 0 \ .
\ee
If we want to compare two gauges, $\chi = \sqrt 6$ and $a = 1$, it is natural to replace $\chi$ by the field $\tilde\chi = \chi/\sqrt 6$, which satisfies equation
\be\label{lambdaWDWconf1}
\left[{d^{2}\over d\tilde\chi^{2}} + 108 \lambda \tilde\chi^{4}\right]\Psi(\tilde\chi) = 0 \ ,
\ee
which has exactly the same form as \rf{lambdaWDW}. What is different here is the interpretation of the cosmological evolution. In the gauge $\chi = \sqrt 6$ (i.e. $\tilde\chi = 1$), the Planck mass is constant and the universe expands exponentially. In the gauge $a = 1$, the universe has a constant size, but the effective Planck mass, proportional to $\chi$, exponentially grows, as discussed in Section \ref{toy}. 

One can easily generalize the results obtained above. For example, for a closed universe, the corresponding Wheeler-DeWitt equation is 
\be
\left[{1\over 24\pi^{2}} {d^{2}\over d \tilde\chi^{2}} - 6\pi^{2} \tilde\chi^{2} + 18\lambda \pi^{2} \tilde\chi^{4}\right] \Psi(\tilde\chi) = 0 \ ,
\ee
which coincides with equation \rf{WDWa} up to the change of variables $\tilde\chi \to a$.

The possibility of a dual description of the cosmological evolution is not just a specific property of a narrow class of theories, such as our toy model \rf{toy2a}. As we will see now, all models of $N = 1$ supergravity in their superconformal formulation share this important property. 

\section{Cosmological evolution in  a generic superconformal theory} 

\subsection{Superconformal theory: A brief reminder 
}
A superconformal theory underlying generic supergravity has an extra scalar multiplet, conformon. This supermultiplet was first introduced in \cite{Siegel:1977hn,Kaku:1978ea}.
The scalar-gravity part of superconformal action in the form given in \cite{Kallosh:2000ve} is based on earlier work  \cite{Cremmer:1978hn}. The detailed information can be found in the textbook\cite{Freedman:2012zz}.  
  is
\be
\mathcal{L}_{\rm conf}= \sqrt{-g}\Big ( -\ft16{\cal N} (X,\bar X)R
-G_{I\bar J}\partial ^\mu X^I\, \partial _\mu \bar X^{\bar J}-V(X, \bar X)\Big ) \, , \qquad I, \bar I=0, 1,...,n.
 \label{cGrav}
\end{equation}
This action is invariant with respect to a local conformal symmetry 
\be
(X^I)'= \rme^{\sigma(x)}  X^I\, , \qquad (\bar X^I)'= \rme^{\sigma(x)}  \bar X^I \, , \qquad 
g_{\mu\nu} '= \rme^{-2\sigma(x)} g_{\mu\nu}\,   \ ,
\label{dilat}\ee
under condition that 
 $\mathcal{N}\left( X, \bar X\right)$ is homogeneous of first degree in both $X$ and $\bar X$. The potential  $V$ is homogeneous of degree 2  in both $X$ and $\bar X$. Thus the scalars have  conformal weight $w=1$ whereas the metric $g_{\mu\nu}$ has $w=-2$ and $g^{\mu\nu}$ has $w=2$.

 The $n+1$ scalars including the compensator multiplet form an embedding  K{\"a}hler manifold with
metric, connection and curvature given, respectively,  by
\be
  G_{I\bar J}=\partial_I \partial_{\bar J} {\cal N} \equiv {\partial  \mathcal{N}(X, \bar X)\over \partial X^I \partial \bar X^{\bar J}} \ ,
\label{LargeK}  \ee
\be \Gamma^I_{JK}= G^{I\bar L}{\cal N}_{JK\bar L} \ ,
\qquad 
  R_{I\bar K J\bar L}= {\cal N}_{IJ\bar K\bar L}-{\cal N}_{IJ\bar M}G^{M\bar M}{\cal N}_{M\bar K\bar L} \ ,
  \label{curv}\ee
where 
\be
{\cal N}_{JK\bar L}=\frac{\partial ^{3}\mathcal{N}}{\partial
X^{J} \partial
X^{K}\partial \overline{X}^{\overline{L}}} \, , \qquad {\cal N}_{IJ\bar K\bar L}= \frac{\partial ^{4}\mathcal{N}}{\partial
X^{I} \partial
X^{J}\partial \overline{X}^{\overline{K}} \partial \overline{X}^{\overline{L}}} \ .
\label{curv1}\ee

In supersymmetric case the potential depends on a superpotential and the scalar derivatives involve the gauge-field of the local $\mathcal{R}$-symmetry.
\be
V(X, \bar X)\Rightarrow  G_{I\bar J} {\cal W}_I \bar {\cal W}_{\bar J}\, , \qquad \partial_\mu \Rightarrow D_\mu= \partial_\mu +i A_\mu \ .
\ee
However, for investigation of the cosmological evolution it is often sufficient to use the locally conformal part of the symmetry, without the requirement of supersymmetry and local $\mathcal{R}$-symmetry.  One example of such models was presented in \cite{Kallosh:2013pby}. We therefore continue our discussion  for locally conformal models, the generalization to the bosonic part of the superconformal ones is straightforward.

\subsection{Einstein frame conformal gauge}

We will start our analysis in the Einstein frame, which means $M_{p} = 1$. In order to do it, one may pick up a conformal gauge which breaks conformal symmetry as follows:
\be
{\cal N} (X,\bar X)=-3\, \ .
\label{einst}\ee
 In this gauge the action is
 \be
\mathcal{L}_{\rm conf}= \sqrt{-g}\Big ( \ft12 R
-G_{I\bar J}\partial ^\mu X^I\,\partial _\mu \bar X^{\bar J}-V(X, \bar X)\Big ) \, , \qquad I, \bar I=0, 1,...,n.
 \label{EGrav}
\end{equation}
where eq. \rf{einst} can be solved in terms of $n$ physical complex scalars $z^i, \bar z^{\bar i}$, where $i, \bar i=1,..., n$. This procedure also requires, in general, to gauge-fix  also the local $U(1)$ $\mathcal{R}$-symmetry, as it was done in many examples of derivation of supergravity from the superconformal theory, see \cite{Freedman:2012zz}.
Thus,  $X, \bar X$ become functions of physical scalars $z^i, \bar z^{\bar i}$.  The remaining action describes the scalars in gravitational field with the Einstein action for gravity, which corresponds to the  Einstein frame. 

A nice example of such a gauge is the `rapidity gauge' used in \cite{Kallosh:2013hoa}, where $\chi^2- \phi^2 =6$.
This condition is resolved so that $\chi= \sqrt 6 \cosh {\varphi\over  \sqrt 6}$ and $\phi= \sqrt 6 \sinh {\varphi\over  \sqrt 6}$, where $\varphi$ is a canonical $z$ field in the Einstein frame.

\subsection{Jordan frame  conformal gauge}  

One may also use a conformal gauge
\be
{\cal N} (X,\bar X)=-3 \, \Phi(z, \bar z) \ ,
\label{Jord}\ee
where $\Phi(z, \bar z)$ is a function of physical scalars specifying the choice of the Jordan frame. In this gauge the action is
 \be
\mathcal{L}_{\rm conf}= \sqrt{-g}\Big ( \ft12 \Phi(z, \bar z) R
-G_{I\bar J}\partial ^\mu X^I\,\partial _\mu \bar X^{\bar J}-V(X, \bar X)\Big ) \, , \qquad I, \bar I=0, 1,...,n \ ,
 \label{JGrav}
\end{equation}
where eq. \rf{Jord} can be solved in terms of $n$ physical scalars, so that $X, \bar X$ are functions of physical scalars $z^i, \bar z^{\bar i}$. The remaining action describes the scalars in gravitational field in Jordan frame  for gravity.
Examples of such gauges are given in the supersymmetric Higgs inflation models in 
\cite{Ferrara:2010yw}, see also \cite{Kallosh:2013pby}.

\subsection {Metric dependent conformal gauge fixing}
Instead of using gauges where some functions of scalars are fixed, which leads to  the Einstein or the Jordan frame actions  above preserving general covariance, one can  fix some combination of the 10 functions in the metric which would break Weyl symmetry. 
We take the following metric \cite{Arnowitt:1959ah,DeWitt:1967yk}:
\be
ds^2=-\alpha^2 dt^2  + \gamma_{ij}( \beta^i dt+dx^i) ( \beta^j dt+dx^j) \ ,
\label{metric}\ee
where  all components of the metric depend on time and space coordinates. The Weyl weight of these functions is the following $w_{\alpha^2}=w_{\gamma_{ij}}=-2$, $w_{\beta^i}=-0$.

We may choose any combination of these 10 functions
with non-vanishing Weyl weight and fix it to a constant.
For example,  we may take
\be
-g\equiv - \det (g_{\mu\nu}) = \alpha^2 \gamma =1 \ ,
\ee
which is particularly useful for 
the \F\, universe with a  conformally flat geometry
$
 ds^2= -a^2(\eta)(d\eta^2 - d\vec x^2)$. In  this gauge $a=1$ and the metric is flat, $
 ds^2=  -d\eta^2 + d\vec x^2$.

\subsection{Conformally covariant evolution}\label{SCEV}

When solving non-linear classical equations of motion in general relativity, there is no need to make a particular choice of a gauge which fixes  reparametrization symmetry. One can use the Einstein and Klein-Gordon equations and solve them by making a particular ansatz for the metric. 

An analogous strategy can be used in case of the general covariance  and Weyl symmetry. Starting from the action \rf{cGrav} we may use a generic ansatz for the metric \rf{metric} and derive all classical equations of motion.  We use  notation in  \cite{DeWitt:1967yk}, where
$
K\equiv \gamma^{ij} K_{ij} $ $ K^{ij} \equiv \gamma^{ik} \gamma^{jl} K_{kl}$ $  K_{ij}\equiv {1\over 2} \alpha^{-1} (\beta_{i,j} + \beta_{j,i} -\gamma_{ij, 0})
$. The first  term in the action \rf{cGrav},  given by $-\ft16{\cal N} (X,\bar X)R$,   becomes
\be
-{1\over 6} \sqrt {-g} \, {\cal N}(X, \bar X) [ \alpha \gamma^{1/2} (K_{ij} K^{ij} - K^{(2)} + \, ^{(3)}R - 2\gamma^{1/2} ( K)_{, 0}+(K\beta^i - \gamma^{ij} \alpha_{,j})_{, j}].
\label{action}\ee
Here $K_{ij}$ is the second fundamental form, $^{(3)}R$ is the intrinsic and $(K_{ij} K^{ij} - K^{(2)})$ is the extrinsic curvature, respectively. The last 3 terms in \rf{action}, being total derivatives in the Einstein frame where $\cal N$ is a constant, will drop from the action. However, in the superconformal theory with a generic Jordan frame function ${\cal N}(X, \bar X)$, the derivatives hit the scalar dependent $\cal N$ and contribute to the action as well as to equations of motion.
The total superconformal action \rf{cGrav} becomes a functional of $X^I(x) , \bar X^{\bar J}(x), \alpha(x), \beta_i(x), \gamma_{ij}(x)$
and their first derivatives, upon integration by parts: 
\be
\sqrt{-g}\Big ( -\ft16{\cal N} (X,\bar X)R
-G_{I\bar J}\partial ^\mu X^I\, \partial _\mu \bar X^{\bar J}-V(X, \bar X)\Big )     \rightarrow   {\cal L} \Big (X^I(x) , \bar X^{\bar J}(x), \alpha(x), \beta_i(x), \gamma_{ij}(x)\Big ).
 \label{ansztzGrav}
\end{equation}
Generic equations of motion can be derived and studied. Since the action \rf{ansztzGrav} still has an unbroken Weyl symmetry, the field equations are covariant under conformal transformation. This means that there is a relation between the solutions of equations for scalars and for the metric, as always in case of gauge symmetries,
\be
{\delta S\over \delta X^I } X^I
 +{\delta S\over \delta \bar X^{\bar J}}  \bar X^{\bar J}
 -2 {\delta S\over \delta g_{\mu\nu}} g_{\mu\nu} =0 \ . 
\ee

These solutions may be compared with those where different choices of the conformal gauges were made from the very beginning, for example leading to the Einstein frame where only physical scalars are left, or to a certain Jordan frame with non-minimally coupled physical scalars. Another possibility is to use the metric-dependent conformal gauge where one of the functions in the metric is fixed, for example the gauge 
$-g\equiv - \det (g_{\mu\nu}) = \alpha^2 \gamma =1$.

Once generic equations of motion following from \rf{ansztzGrav} have been solved, one may start addressing the following issue.
Solutions in any conformal gauge will define solutions in any other conformal gauge due to conformal symmetry  of the action in \rf{ansztzGrav}.
Is it possible to use the advantages of some of the conformal  gauges over the other? The answer is positive as we have already explained in simple examples in Section 2. Below we will consider two very different stages of the cosmological evolution, one is the inflationary period where the initial deviations from the FLRW metric decrease, so that one can use advantages of the conformal flatness of the Friedmann universe, and an opposite regime when we study an approach to a cosmological singularity where the fate of initial deviations from the FLRW metric is an issue.

\section{Conformally flat classical FLRW geometry}

\subsection{Geodesic equations for scalar fields in moduli space in a gauge $a(\eta) = 1$}

During inflation, the conformally flat FLRW metric is an attractor solution of equations of motion forward in time: the initial anisotropy and inhomogeneity  decrease exponentially, and the metric rapidly approaches the one of a flat Friedmann universe (up to small perturbations due to quantum effects),
$
 ds^2= -a^2(\eta)(d\eta^2 - d\vec x^2)
$
The action \rf{cGrav} has a local conformal symmetry which allows us to choose a conformal gauge
$
a(\eta) =1$. 

Equations which define the cosmological evolution in such case follow from the  action 
 \be
\mathcal{L}_{\rm flat}= 
-G_{I\bar J}\partial ^\mu X^I\,\partial _\mu \bar X^{\bar J}-V(X, \bar X) \, , \qquad I, \bar I=0, 1,...,n.
 \label{cGravFlat}
\end{equation}
since  $R=0$. 
Equations of motion for $n+1$ scalars are {\it geodesic equations in the moduli space geometry affected  by the potential},
\be 
{\partial ^2 X^I
\over \partial \eta^2} + \Gamma^I_{JK} {\partial  X^J
\over \partial \eta}  {\partial  X^K
\over \partial \eta} + G^{I\bar J} {\partial V\over \partial \bar X^{\bar j}}=0 \ .
\label{new}\ee
Here $n+1$ scalars obey equations of motion in a flat Minkowski space. They are coupled due to internal moduli space geometry and via potential term in the action.

The conformal symmetry of the action means that the solutions are the same (related by the change of variables) as the ones which could be obtained using the standard equations of motion in general relativity for the scale factor $a(\tau)$ and for the $n$ scalars in the gravitational field. The standard equations including the Friedmann equations plus the $n$ Klein-Gordon equations in gravitational field  here are  replaced by  $n+1$ eqs. \rf{new} for scalars interacting in a flat space.
Many  examples can be studied.

\subsection{Hamiltonian formalism}
Here we start with a slightly more general metric including the lapse function, as we did in section 3. We assume that fields depend only on time. In $a=1$ gauge the action  is 
\be
\mathcal{L}= 
{1\over N(t)}G_{I\bar J}\dot X^I\,\dot {\bar X}^{\bar J}- N(t)V(X, \bar X) \, , \qquad I, \bar I=0, 1,...,n.
 \label{cGravFlat1}
\end{equation}
The action does not depend on the lapse velocity $\dot N$. Therefore the Lagrange equation for $N(t)$ is
\be
{\partial {\mathcal L}\over \partial N(t)}= -{1\over  N^2} G_{I\bar J}\dot X^I\,\dot {\bar X}^{\bar J}   -V(X, \bar X)=0\, \ .
\label{Neq}\ee
For the choice of a conformal time variable $\eta$ with $N(t)=1$  with $d \eta= N(t) dt$ we find a constraint
\be
G_{I\bar J}( X^I)' \,( {\bar X}^{\bar J})'   +V(X, \bar X)=0\, ,
\ee
where $'$ is the derivative over $\eta$.
We define the canonical momenta
\be
\pi_I= {\partial \mathcal{L}\over \partial \dot X^I}= {1\over N(t)} G_{I\bar J} \cdot \dot {\bar X}^{\bar J} =  G_{I\bar J} \cdot ( {\bar X}^{\bar J})'\ , \qquad \pi_N={\partial \mathcal{L}\over \partial \dot N}=0 \ ,
\ee
and  
 \be
\pi_N \dot N + \pi_I\dot X^I    -{\cal L} = \pi_N \dot N + N {\cal H} \ ,
\ee
where 
\be
\mathcal{H}= \pi_I G^{I\bar J} (X, \bar X) \bar \pi_{\bar J} + V(X, \bar X) \ .
\ee
The primary constraint $\pi_N=0 $ is associated with the { secondary or dynamical constraint}, the Hamiltonian constraint 
\be
\mathcal{H}=\pi_I G^{I\bar J} (X, \bar X) \bar \pi_{\bar J} + V(X, \bar X) =0 \ ,
\ee 
as in the simple example  studied in section 3. This constraint coincides with the equation of motion of $N(t)$ in \rf{Neq} since it states that
\be
\dot \pi_N= {\partial {\mathcal L}\over \partial N(t)}=0 \qquad \Rightarrow \qquad \mathcal{H}=0 \ .
\ee

Note that our moduli space metric $G_{I\bar J}$ is not positive definite, the conformon has a negative signature $G_{0\bar 0}=-1$, whereas the physical fields have a positive signature. Therefore the vanishing of the Hamiltonian means that the kinetic and potential energy from the conformon and from the physical scalars compensate each other.

In supersymmetric case everything above is valid for the special case when the potential depends on the superpotential
\be
\mathcal{H}= \pi_I G^{I\bar J}  \bar \pi_{\bar J} + \mathcal{W}_I G^{I\bar J}  \bar {\mathcal{W}}_{\bar J}=0 \ .
\label{susy}\ee

\section{Towards the cosmological singularity with conformal symmetry}
If we are interested in the cosmological evolution towards the cosmological singularity, one may ask the following question. In general relativity an obvious signal of the cosmological singularity is the vanishing of the scale factor $a(t_{\rm sing})=0$ at the time of singularity at $t=t_{\rm sing}$. However, this kind of a signal may be misleading, as evidenced by the investigation of the event horizon surrounding the black hole, where $g_{tt} =0$ but this is just a coordinate singularity, which may be removed by the change of coordinates. 

According to  \cite{Landau:1989gn}, cosmological singularity is a place where the density of matter and the invariants of the curvature tensor become infinite.  There are several different reasons why this definition of the singularity makes sense and is widely accepted in the cosmological literature. First of all, the singularity of the curvature invariants such as $R_{\mu\nu\lambda\delta}R^{\mu\nu\lambda\delta}$ or the Weyl tensor squared $C_{\mu\nu\lambda\delta}C^{\mu\nu\lambda\delta}$ cannot be removed by the change of coordinates, unlike the event horizon singularity of the metric near the black holes. Also, terms like that can appear as higher order corrections in the effective action of general relativity. The general structure of the effective Lagrangian in GR with such corrections can be schematically represented as 
\be\label{eff}
L_{\rm eff} =\sqrt{-g}\, \left( {M_{p}^{2}\over 2}\, R+c_2 R^2 +c_3{R^3\over M_{p}^{2}}+...\right ),
\ee
where we suppressed all indices, and omitted terms of more complicated structure which may be present in this expression. When the curvature invariants such as $R_{\mu\nu\lambda\delta}R^{\mu\nu\lambda\delta}$ become greater than $M_{p}^{4} = 1$, higher order terms in the effective action become more important than the lower order terms, assuming that the coefficients $c_{n}$ do not become vanishingly small for large $n$. This is closely related to the common lore that the standard methods of quantum gravity and supergravity are expected to fail at super-Planckian energies, which is a manifestation of the UV incompleteness of quantum gravity. One may try to address this problem using methods of string theory, but the progress in investigation of the cosmological singularities by such methods is still rather limited.

In this section we will describe an approach to investigation of cosmological singularities based on the hidden superconformal symmetry discussed in this paper. In this approach, the standard Einstein theory corresponds to a certain choice of a gauge. % so one may wonder whether the requirement of gauge invariance may constrain the structure of higher order terms of the type discussed above. 
 One may wonder whether some other gauges are better suited for investigation of the cosmological singularities. %(just like the renormalizable gauges are better suited for investigation of UV divergences in the standard model than the unitary gauge). 
 If in some of these gauges the singularities disappear, % and the curvature invariants never approach the Planck density, 
then one could use such gauges for a reliable description of the cosmological evolution without worrying too much about the UV completion of the theory.

To analyze this possibility, one should note that any solution of equations of motion described in section \ref{SCEV} on `Conformally covariant evolution' with an arbitrary ansatz for the metric transforms covariantly under Weyl transformations, since both the metric and scalars transform. Generic solutions found in any given gauge are related to solutions in other gauges, such as the Einstein frame gauge, the Jordan frame gauge, or $g=-1 $ gauge, by some Weyl transformations. 

Our goal is to construct invariants depending on the curvature tensor, which are invariant under the change of coordinates
\be
x^\mu \rightarrow x^\mu + \xi^\mu(x)\, , 
\label{GR}\ee
where $\xi^\mu(x) $ is an arbitrary function of coordinates,  and under  Weyl  conformal gauge transformations  where the metric has $w=-2$ and the scalars have $w=1$,  
\be
 g_{\mu\nu} \rightarrow e^{ -2\sigma(x)} g_{\mu\nu}\, , \qquad  \delta X \rightarrow e^{ \sigma(x) } X \ .
\label{dilatsmall}
\ee

Thus, we are looking for a generalization of the invariants like $R_{\mu\nu\lambda\delta}R^{\mu\nu\lambda\delta}$ and analogous higher order scalars build from the Riemann tensor $R_{\mu\nu\lambda\delta}$. The 4-tensor which is invariant under Weyl transformation of the metric \rf{dilat} is a Weyl tensor, which is a traceless part of the Riemann tensor
\be
C^\mu{} _{\mu\lambda\delta}, \qquad w=0 \ .
\ee
To make the simplest scalar out of the Weyl invariant tensor one should make a contraction of two such tensors with two inverse metric tensors, which will produce a reparametrization scalar 
\be
C_{\mu\nu \lambda\delta} C^{\mu\nu \lambda\delta} , \qquad w=4 \ .
\ee
 To make this product Weyl invariant,
 we need 
 to find another Weyl covariant scalar with the Weyl weight $-4$. The simplest possibility is to use our \K\, potential of the embedding space, ${\cal N}(X, \bar X)$,   with the Weyl weight  $w=2$. Thus, the  simplest   Weyl invariant which is coordinate independent under \rf{GR} is given by the square of the Weyl tensor weighted by the inverse \K\, potential of the embedding space,
 \be
 I= \Big [{3 \over {\cal N}(X, \bar X) }\Big]^{2}\, C_{\mu\nu \lambda\delta} C^{\mu\nu \lambda\delta} \ .
\label{I} \ee

In the Einstein frame gauge, where $ {\cal N}(X, \bar X) =-3$, this invariant reduces to the square of the Weyl tensor
 \be
 I_E=  (C_{\mu\nu \lambda\delta} C^{\mu\nu \lambda\delta})_E \ .
 \ee
% Restoring the Planck mass in this equation, we find that the invariant \rf{I} generalizes the term
% \be
% I_E=  {(C_{\mu\nu \lambda\delta} C^{\mu\nu \lambda\delta})_E\over M_{p}^{4}} \ .
% \ee
 In the Jordan frame gauge, where $ {\cal N}(X, \bar X) =-3\, \Phi(z,\bar z)$, it becomes
\be
 I_J=  \Phi(z,\bar z) ^{-2}\, (C_{\mu\nu \lambda\delta} C^{\mu\nu \lambda\delta})_J \ .
 \ee
More generally, if the Weyl symmetry is gauge-fixed in any other way, one should just compute the expression \rf{I} in that particular gauge. By construction, \rf{I} is independent on the choice of the conformal gauge.
 
One can form many other generic functions of scalars $X, \bar X$ with required weight $w$ to compensate for powers of the Weyl tensor.  For example, one can take 
\be
 I= f(X, \bar X)\,  C_{\mu\nu \lambda\delta} C^{\mu\nu \lambda\delta} \ ,
\label{I1} \ee 
where $f(X, \bar X)$ is any function with $w=-4$.

One can easily construct higher order Weyl invariant scalars, e.g.
\be
 \Big [{3 \over {\cal N}(X, \bar X) }\Big]^{3} \, C_{\mu\nu}{}^{\lambda\delta} C_{\lambda\delta}{}^{\alpha \beta}   C_{\alpha\beta}{}^{\mu\nu} \ .
\label{prediction}
 \ee
 
 To summarize, when equations of motion for the Weyl invariant models are solved, one can study some properties of these solutions which are specific for a given choice of the conformal gauge. However, there are geometric invariants, such as \rf{I} and \rf{prediction}, which are independent on such choices and also are reparametrization scalars. Only such invariants may serve as observables representing those features of the solutions that are gauge independent as well as independent on the choice of a coordinate system.

One may also look at Weyl invariants which are not scalars, but transform as  densities, for example
\be\label{weight}
\tilde I= \sqrt {-g} \, C_{\mu\nu \lambda\delta} C^{\mu\nu \lambda\delta} \ ,
\ee
since $\sqrt {-g}$ has weight $w=-4$. The higher order terms of this type include
\be
 \sqrt {-g} \Big [{3 \over {\cal N}(X, \bar X) }\Big] \, C_{\mu\nu}{}^{\lambda\delta} C_{\lambda\delta}{}^{\alpha \beta}   C_{\alpha\beta}{}^{\mu\nu} \ .
\label{prediction1}
 \ee

These scalar densities  are Weyl invariant, so one can add them to the effective Lagrangian  just as we did with the higher order terms in equation \rf{eff}. By comparing \rf{weight} and \rf{prediction1}, one finds that these invariants, just like the higher order terms in \rf{eff}, contain growing powers of ${C_{\mu\nu \lambda\delta}/{\cal N}(X, \bar X)}$. Therefore one may expect that the higher order corrections to equations of motion generically become greater than the classical part of the equations  when the invariants of the type of \rf{I} and \rf{prediction} become greater than $O(1)$. This provides the Weyl invariant generalization of the concept of the Planck curvature/density in the Einstein gauge. Finally, if the invariants such as \rf{I} and \rf{prediction} diverge, one has a cosmological singularity. If these invariants are singular in the Einstein gauge, the same singularity appears in all other gauges.

\section{Discussion}

The latest developments in inflationary cosmology suggest that the superconformal formalism is much more than  just a powerful tool for the development of supergravity. The superconformal formulation of supergravity possessed several additional symmetries, including local conformal symmetry (Weyl symmetry). Unification of this formulation and inflationary cosmology may lead to many interesting implications. Indeed, inflation generically makes the universe homogeneous and isotropic, so one can describe it by the FLRW metric, which is conformally flat. This may allow various simplifications in the investigation of physical processes in the early universe. 

Perhaps more significantly, additional symmetries which are present in the superconformal formulation provide a natural framework for formulating new classes of cosmological models, which would be very difficult to construct in the more familiar framework of the Poincar\'e supergravity where
the conformal compensator field (which we will call conformon) is fixed. As a most recent example, we would mention a broad class of  inflationary models based on spontaneously broken conformal or superconformal symmetry  \cite{Kallosh:2013pby,Kallosh:2013lkr,Kallosh:2013hoa,Ferrara:2013rsa,Kallosh:2013daa,Kallosh:2013tua}. These models look very natural in the original superconformal formulation, they lead to universal model-independent predictions in excellent agreement with the recent cosmological data from WMAP9  \cite{Hinshaw:2012aka} and Planck 2013 \cite{Ade:2013rta}, but it would be really hard to identify such theories without using the superconformal approach.

Therefore in this paper we continued developing the superconformal approach in application to cosmology. In particular, we found that the cosmological expansion in the Friedmann universe can be equivalently described as a growth of the conformon field in the non-expanding Minkowski space; the cosmological inflation has a dual description as the conformon instability. We derived the Wheeler-DeWitt equations for the wave function of the universe depending on the conformon field instead of the scale factor of the universe. Thus the geometry of our world can be equally well described in terms of the scale factor of the universe or in terms of the conformal compensator.
But it is not just ``either or'' approach: We developed a full geometric formulation describing a simultaneous evolution of the metric and scalar fields, including the conformal compensator. This extends the superspace approach initiated by Bryce DeWitt \cite{DeWitt:1967yk}, and provides a generalized interpretation of the famous Hamiltonian constraint which implies that the total Hamiltonian of the universe vanishes.

Whereas one is free to use any gauge describing the superconformal evolution, it is important to have a set of invariant quantities which do not depend on the choice of the gauge. In this paper we constructed a set of invariants, generalizing the invariants of the type of $R_{\mu\nu \lambda\delta} R^{\mu\nu \lambda\delta}$ or $C_{\mu\nu \lambda\delta} C^{\mu\nu \lambda\delta}$, which are traditionally used for studies of space-time singularities in GR. The simplest of these invariants is given by a combination of the square of the Weyl curvature tensor and the inverse square of the \K\, potential of the embedding manifold, $I= \Big [{3 \over {\cal N}(X, \bar X)  }\Big]^{2} C_{\mu\nu \lambda\delta} C^{\mu\nu \lambda\delta}$.
Since  ${\cal N} (X,\bar X)=-3$ in the Einstein frame, this result implies that if any cosmological solution has a singularity in the Einstein frame, such that $(C_{\mu\nu \lambda\delta} C^{\mu\nu \lambda\delta})_E\rightarrow \infty $, this solution remains singular  in an arbitrary Weyl transformed geometry: the value of the Weyl invariant $I$ \rf{I} will be the same as the one in the Einstein frame. A more detailed  discussion of this issue and its cosmological implications  will be contained in a separate publication \cite{Carrasco:2013hua}.

\section*{Acknowledgments}

This article was strongly influenced by a continuous collaboration and exchange of ideas on the superconformal approach to cosmology with Lev Kofman.  We are grateful to S. Ferrara, A. Marrani, M. Porrati,   D. Roest  and A.~Van~Proeyen for collaboration on some of the papers which influenced this project, and to J. J. M. Carrasco and W. Chemissany for many stimulating discussions. This work was supported by the SITP and by the NSF Grant PHY-1316699. The work of RK was also supported by the Templeton Foundation Grant ``Frontiers of Quantum Gravity''.


\begin{thebibliography}{99}

   
 %\cite{Siegel:1977hn}
\bibitem{Siegel:1977hn} 
  W.~Siegel,
 ``A Polynomial Action for a Massive, Self-interacting Chiral Superfield Coupled to Supergravity,''
  HUTP-77/A077.
  %%CITATION = HUTP-77/A077;%%
%\cite{Gates:1983nr}
%\bibitem{Gates:1983nr} 
  S.~J.~Gates, M.~T.~Grisaru, M.~Rocek and W.~Siegel,
  ``Superspace Or One Thousand and One Lessons in Supersymmetry,''
  Front.\ Phys.\  {\bf 58}, 1 (1983)
  [hep-th/0108200].
  %%CITATION = HEP-TH/0108200;%%
  %382 citations counted in INSPIRE as of 23 Apr 2013
  
 %\cite{Kaku:1978ea}
\bibitem{Kaku:1978ea} 
  M.~Kaku and P.~K.~Townsend,
  ``Poincare Supergravity As Broken Superconformal Gravity,''
  Phys.\ Lett.\ B {\bf 76}, 54 (1978).
  %%CITATION = PHLTA,B76,54;%%
  %97 citations counted in INSPIRE as of 23 Apr 2013 
  



  
  \bibitem{Cremmer:1978hn}  E.~Cremmer, B.~Julia, J.~Scherk, S.~Ferrara,
L.~Girardello and P.~van Nieuwenhuizen, ``Spontaneous Symmetry
Breaking And Higgs Effect In Supergravity Without Cosmological
Constant,'' Nucl.\ Phys.\
 B {\bf 147}, 105 (1979).
   %\bibitem{BFNS-82}  R.~Barbieri, S.~Ferrara, D.~V.~Nanopoulos and K.~S.~Stelle,
``Supergravity, R Invariance And Spontaneous Supersymmetry Breaking,''
  Phys.\ Lett.\  B {\bf 113}, 219 (1982).
   %\cite{Kugo:1982cu}
%\bibitem{Kugo:1982cu} 
  T.~Kugo and S.~Uehara,
  ``Conformal And Poincare Tensor Calculi In N=1 Supergravity,''
  Nucl.\ Phys.\ B {\bf 226}, 49 (1983).
  %%CITATION = NUPHA,B226,49;%%
  %117 citations counted in INSPIRE as of 25 May 2013
%\bibitem{CFGVP-1} 
 E.~Cremmer, S.~Ferrara, L.~Girardello and A.~Van
Proeyen, ``Yang-Mills Theories With Local Supersymmetry:
Lagrangian, Transformation Laws And Superhiggs Effect,'' Nucl.\ Phys.\
B {\bf 212}, 413 (1983).


%\cite{Freedman:2012zz}
\bibitem{Freedman:2012zz} 
  D.~Z.~Freedman and A.~Van Proeyen,
  ``Supergravity,''
  Cambridge, UK: Cambridge Univ. Pr. (2012) 607 p




%\cite{Kallosh:2000ve}
\bibitem{Kallosh:2000ve}
  R.~Kallosh, L.~Kofman, A.~D.~Linde and A.~Van Proeyen,
``Superconformal symmetry, supergravity and cosmology,''
Class.\ Quant.\ Grav.\  {\bf 17}, 4269 (2000)
 [Erratum-ibid.\  {\bf 21}, 5017 (2004)]
 [arXiv:hep-th/0006179].
  %%CITATION = CQGRD,17,4269;%
  
 %\cite{Ferrara:2010yw}
\bibitem{Ferrara:2010yw}
  S.~Ferrara, R.~Kallosh, A.~Linde, A.~Marrani and A.~Van Proeyen,
``Jordan Frame Supergravity and Inflation in NMSSM,'' Phys. Rev. {\bf D82}, 045003
(2010)
  [arXiv:1004.0712 [hep-th]].
  %%CITATION = ARXIV:1004.0712;%%
  %\cite{Ferrara:2010in}
%\bibitem{Ferrara:2010in} 
  S.~Ferrara, R.~Kallosh, A.~Linde, A.~Marrani, A.~Van Proeyen and ,
 ``Superconformal Symmetry, NMSSM, and Inflation,''
  Phys.\ Rev.\ D {\bf 83}, 025008 (2011)
  [arXiv:1008.2942 [hep-th]].
  %%CITATION = ARXIV:1008.2942;%%
  %50 citations counted in INSPIRE as of 30 Mar 2013
  
  %\cite{Kallosh:2013pby}
\bibitem{Kallosh:2013pby} 
  R.~Kallosh and A.~Linde,
  ``Superconformal generalization of the chaotic inflation model $\frac{\lambda}{4} \phi^{4} - \frac{\xi}{2} \phi^{2}R$,''
  JCAP {\bf 1306}, 027 (2013)
  [arXiv:1306.3211 [hep-th]].
  %%CITATION = ARXIV:1306.3211;%%
  %2 citations counted in INSPIRE as of 02 Jul 2013
  
   %\cite{Kallosh:2013lkr}
\bibitem{Kallosh:2013lkr} 
  R.~Kallosh and A.~Linde,
 ``Superconformal generalizations of the Starobinsky model,''
  JCAP {\bf 1306}, 028 (2013)
  [arXiv:1306.3214 [hep-th]].
  %%CITATION = ARXIV:1306.3214;%%
  %3 citations counted in INSPIRE as of 02 Jul 2013
  
 %\cite{Kallosh:2013hoa}
\bibitem{Kallosh:2013hoa} 
  R.~Kallosh and A.~Linde,
``Universality Class in Conformal Inflation,''
  JCAP {\bf 1307}, 002 (2013)
  [arXiv:1306.5220 [hep-th]].
  
 %\cite{Ferrara:2013rsa}
\bibitem{Ferrara:2013rsa} 
  S.~Ferrara, R.~Kallosh, A.~Linde and M.~Porrati,
``Minimal Supergravity Models of Inflation,''
Phys. Rev. D {\bf 88}, 085038 (2013)
  [arXiv:1307.7696 [hep-th]].
  %%CITATION = ARXIV:1307.7696;%%
  %15 citations counted in INSPIRE as of 06 Nov 2013 
  
  %\cite{Kallosh:2013daa}
\bibitem{Kallosh:2013daa} 
  R.~Kallosh and A.~Linde,
``Multi-field Conformal Cosmological Attractors,''
JCAP {\bf 1312}, 006 (2013)
  [arXiv:1309.2015 [hep-th]].
  %%CITATION = ARXIV:1309.2015;%%
  %4 citations counted in INSPIRE as of 06 Nov 2013

%\cite{Kallosh:2013tua}
\bibitem{Kallosh:2013tua} 
  R.~Kallosh, A.~Linde and D.~Roest,
``A universal attractor for inflation at strong coupling,''
Phys. Rev. Lett.  {\bf 112}, 011303 (2014)
  arXiv:1310.3950 [hep-th].
  %%CITATION = ARXIV:1310.3950;%%
  %2 citations counted in INSPIRE as of 06 Nov 2013
  
%\cite{Kallosh:2013yoa}
\bibitem{Kallosh:2013yoa} 
  R.~Kallosh, A.~Linde and D.~Roest,
``Superconformal Inflationary $\alpha$-Attractors,''
  JHEP {\bf 1311}, 198 (2013)
  [arXiv:1311.0472 [hep-th]].
  %%CITATION = ARXIV:1311.0472;%%
    
%\cite{Hinshaw:2012aka}
\bibitem{Hinshaw:2012aka} 
  G.~Hinshaw {\it et al.}  [WMAP Collaboration],
 ``Nine-Year Wilkinson Microwave Anisotropy Probe (WMAP) Observations: Cosmological Parameter Results,''
  Astrophys.\ J.\ Suppl.\  {\bf 208}, 19 (2013)
  [arXiv:1212.5226 [astro-ph.CO]].
  %%CITATION = ARXIV:1212.5226;%%
  %628 citations counted in INSPIRE as of 07 Jan 2014

%\cite{Ade:2013rta}
\bibitem{Ade:2013rta} 
  P.~A.~R.~Ade {\it et al.}  [ Planck Collaboration],
``Planck 2013 results. XXII. Constraints on inflation,''
  arXiv:1303.5082 [astro-ph.CO].
  %%CITATION = ARXIV:1303.5082;%%
  %9 citations counted in INSPIRE as of 31 Mar 2013
  P.~A.~R.~Ade {\it et al.}  [Planck Collaboration],
``Planck 2013 results. XVI. Cosmological parameters,''
  arXiv:1303.5076 [astro-ph.CO].
  %%CITATION = ARXIV:1303.5076;%%
  %291 citations counted in INSPIRE as of 25 Jun 2013
  

  
 
%\cite{Arnowitt:1959ah}
\bibitem{Arnowitt:1959ah} 
  R.~L.~Arnowitt, S.~Deser and C.~W.~Misner,
  ``Dynamical Structure and Definition of Energy in General Relativity,''
  Phys.\ Rev.\  {\bf 116}, 1322 (1959).
  %%CITATION = PHRVA,116,1322;%%
  %182 citations counted in INSPIRE as of 01 Jul 2013  
  
  %\cite{DeWitt:1967yk}
\bibitem{DeWitt:1967yk} 
  B.~S.~DeWitt,
  ``Quantum Theory of Gravity. 1. The Canonical Theory,''
  Phys.\ Rev.\  {\bf 160}, 1113 (1967).
  %%CITATION = PHRVA,160,1113;%%
  %1620 citations counted in INSPIRE as of 01 Jul 2013
  
  %\cite{Vilenkin:1982de}
\bibitem{Vilenkin:1982de} 
  A.~Vilenkin,
``Creation of Universes from Nothing,''
  Phys.\ Lett.\ B {\bf 117}, 25 (1982).
  %%CITATION = PHLTA,B117,25;%%
  %385 citations counted in INSPIRE as of 08 Nov 2013
  
%\cite{Hartle:1983ai}
\bibitem{Hartle:1983ai} 
  J.~B.~Hartle and S.~W.~Hawking,
``Wave Function of the Universe,''
  Phys.\ Rev.\ D {\bf 28}, 2960 (1983).  
  
  %\cite{Linde:1983cm}
\bibitem{Linde:1983cm} 
  A.~D.~Linde,
``Quantum creation of an inflationary universe,''
  Sov.\ Phys.\ JETP {\bf 60}, 211 (1984)
  [Zh.\ Eksp.\ Teor.\ Fiz.\  {\bf 87}, 369 (1984)].
  %%CITATION = SPHJA,60,211;%%
  %158 citations counted in INSPIRE as of 08 Nov 2013
A.~D.~Linde,
``Quantum Creation of the Inflationary Universe,''
  Lett.\ Nuovo Cim.\  {\bf 39}, 401 (1984).
  %%CITATION = NCLTA,39,401;%%
  A.~D.~Linde,
``The Inflationary Universe,''
  Rept.\ Prog.\ Phys.\  {\bf 47}, 925 (1984).
  %%CITATION = RPPHA,47,925;%%
  
  %\cite{Rubakov:1984bh}
\bibitem{Rubakov:1984bh} 
  V.~A.~Rubakov,
``Quantum Mechanics in the Tunneling Universe,''
  Phys.\ Lett.\ B {\bf 148}, 280 (1984).
  %%CITATION = PHLTA,B148,280;%%
  
  %\cite{Zeldovich:1984vk}
\bibitem{Zeldovich:1984vk} 
  Y.~B.~Zeldovich and A.~A.~Starobinsky,
``Quantum creation of a universe in a nontrivial topology,''
  Sov.\ Astron.\ Lett.\  {\bf 10}, 135 (1984).
  %%CITATION = SALED,10,135;%%
  
  %\cite{Vilenkin:1984wp}
\bibitem{Vilenkin:1984wp} 
  A.~Vilenkin,
``Quantum Creation of Universes,''
  Phys.\ Rev.\ D {\bf 30}, 509 (1984).
  %%CITATION = PHRVA,D30,509;%%
  %402 citations counted in INSPIRE as of 08 Nov 2013
  
  %\cite{Coule:1999wg}
\bibitem{Coule:1999wg} 
  D.~H.~Coule and J.~Martin,
``Quantum cosmology and open universes,''
  Phys.\ Rev.\ D {\bf 61}, 063501 (2000)
  [gr-qc/9905056].
  
  %\cite{Linde:2004nz}
\bibitem{Linde:2004nz} 
  A.~D.~Linde,
``Creation of a compact topologically nontrivial inflationary universe,''
  JCAP {\bf 0410}, 004 (2004)
  [hep-th/0408164].
  %%CITATION = HEP-TH/0408164;%%

   %\cite{Landau:1987gn}
\bibitem{Landau:1989gn} 
  L.~D.~Landau, and E.~M.~Lifshitz, 
  ``Textbook On Theoretical Physics. Vol. 2: The Classical Theory of Fields,''
  Fourth Revised English Edition,
Pergamon Press (1989) 402p



 %\cite{Carrasco:2013hua}
\bibitem{Carrasco:2013hua} 
  J.~J.~M.~Carrasco, W.~Chemissany and R.~Kallosh,
 ``Journeys Through Antigravity?,''
  arXiv:1311.3671 [hep-th].
  %%CITATION = ARXIV:1311.3671;%%
  %2 citations counted in INSPIRE as of 18 Dec 2013    
\end{thebibliography}
\end{document}